# Market-Based Asset Price Probability

Victor Olkhov

Independent, Moscow, Russia

victor.olkhov@gmail.com

ORCID: 0000-0003-0944-5113

Jan. 20, 2026

**Abstract**

The random values and volumes of consecutive trades made at the exchange with shares of security determine its mean, variance, and higher statistical moments. The volume weighted average price (VWAP) is the simplest example of such a dependence. We derive the dependence of the market-based variance and 3rd statistical moment of prices on the means, variances, covariances, and 3rd moments of the values and volumes of market trades. The usual frequency-based assessments of statistical moments of prices are the limited case of market-based statistical moments if we assume that all volumes of consecutive trades with security are constant during the averaging interval. To forecast market-based variance of price, one should predict the first two statistical moments and the correlation of values and volumes of consecutive trades at the same horizon. We explain how that limits the number of predicted statistical moments of prices by the first two and the accuracy of the forecasts of the price probability by the Gaussian distribution. This limitation also reduces the reliability of Value-at-Risk by Gaussian approximation. The accounting for the randomness of trade volumes and the use of VWAP results in zero price-volume correlations. To study the price-volume empirical statistical dependence, one should calculate correlations of prices and squares of trade volumes or correlations of squares of prices and volumes. To improve the accuracy and reliability of large macroeconomic and market models like those developed by BlackRock's Aladdin, JP Morgan, and the U.S. Fed., the developers should explicitly account for the impact of random trade volumes and use market-based statistical moments of asset prices.

---

This research received no support, specific grant or financial assistance from funding agencies in the public, commercial or nonprofit sectors. We welcome valuable offers of grants, support and positions.

## 1. Introduction

The history of asset pricing research (Dimson and Mussavian, 1999) tracked price probability up to Bernoulli's studies in 1738, but possibly Bachelier (1900) was the first who really highlighted the probabilistic character of price behavior and forecasting. "The probabilistic description of financial prices, pioneered by Bachelier." (Mandelbrot et al., 1997). "In fact, the first author to put forward the idea to use a random walk to describe the evolution of prices was Bachelier." (Shiryaev, 1999). During the last century, countless papers studied models of random prices (Kendall and Hill, 1953; Muth, 1961; Sharpe, 1964; Fama, 1965; Stigler and Kindahl, 1970; Black and Scholes, 1973; Merton, 1973; Tauchen and Pitts, 1983; Mackey, 1989; Friedman, 1990; Cochrane and Hansen, 1992; Campbell, 2000; Heaton and Lucas, 2000; Cochrane, 2001; Poon and Granger, 2003; Andersen et al., 2005; 2006; Cochrane, 2005; Wolfers and Zitzewitz, 2006; DeFusco et al., 2017; Weyl, 2019; Cochrane, 2022). Shiryaev (1999) and Shreve (2004) gave a probabilistic description of prices.

Numerous studies describe the dependence of prices on the market (Fama, 1965; Tauchen and Pitts, 1983; Odean, 1998; Poon and Granger, 2003; DeFusco et al., 2017), on macroeconomics (Cochrane and Hansen, 1992; Heaton and Lucas, 2000; Diebold and Yilmaz, 2008), on business cycles (Mills, 1946; Campbell, 1998), on expectations (Muth, 1961; Campbell and Shiller, 1988; Greenwood and Shleifer, 2014), on trading volumes (Karpoff, 1987; Campbell et al., 1993; Gallant et al., 1992; Brock and LeBaron, 1995; Llorente et al., 2001), and on many other factors that impact price change. The line of factors and references can be continued (Goldsmith and Lipsey, 1963; Andersen et al., 2001; Hördahl and Packer, 2007; Fama and French, 2015).

The conventional description of price probability $P(p)$ is based on the frequency analysis of trades at a price $p$ (Shiryaev, 1999). If $m_p$ is the number of trades at a price $p$ and $N$ is the total number of trades during the averaging interval $\Delta$, then the probability $P(p)$ of a price $p$ is assessed as:

$$P(p) \sim \frac{m_p}{N} \qquad (1.1)$$

$N$ terms of the time series of price $p(t_i)$ during the averaging interval $\Delta$ approximate the *n-th* statistical moment of price $\pi(t;n)=E[p^n(t_i)]$ (1.2):

$$\pi(t;n) = E[\,p^n(t_i)] \sim \frac{1}{N}\sum_{i=1}^{N} p^n(t_i) \quad ; \quad n = 1,2,.. \qquad (1.2)$$

In this paper we study the time series of the values and volumes of consecutive trades made at the exchange with shares of a security during the averaging interval $\Delta$. All factors that impact trading decisions are already accounted for in the time series of the performed market trades.

We study statistical properties of the time series of the trades that were already made in the market. That allows us to ignore any complexities related to the agents' expectations, market shocks, and any risks that may impact agents' trade decisions.

At current time *t,* let us select an averaging interval $\Delta$ (1.3) and consider $N$ terms of the time series of successive trades made at the exchange during $\Delta$. At modern exchanges, consecutive trades are made with a short time span between the trades $\varepsilon<<\Delta$, and any averaging interval $\Delta$ (1.3) contains only a finite number $N$ of trades.

$$\Delta= \left[t-\frac{\Delta}{2};t+\frac{\Delta}{2}\right] \quad ; \quad t_i \in \Delta \quad ; \quad t_{i+1}=t_i+\varepsilon \quad ; \quad i=1,2,..N \tag{1.3}$$

At modern exchange the period $\varepsilon$ between consecutive trades may be less than a second. The values and volumes of consecutive trades at the modern exchange are severely irregular or random. The collecting and processing of market data of consecutive trades with high frequency is rather difficult and not too useful for the description of the mean and variance of price averaged during hours, days, or weeks. To overcome these challenges, one should sum the initial high-frequency time series of values and volumes of consecutive trades during period $\varepsilon_c$ that may be equal to minutes, hours, or days. The duration of the period $\varepsilon_c$ determines the intervals between consecutive trades, and one should choose it according to the problem under consideration. If one is looking for the mean and variance of price that are averaged during interval $\Delta$ equal to 1 hour or 1 day, one should select a period $\varepsilon_c$ to obtain a sufficient number $N>>1$ of terms of consecutive trades during $\Delta$, so $\varepsilon_c<<\Delta$. One may consider that $N\sim 50-100$ terms may be sufficient to derive reasonable approximations of mathematical expectations.

Let us denote the values $C(t_i)$ and volumes $U(t_i)$ of consecutive trades at time $t_i$ that were made during $\Delta$ (1.3) and define prices $p(t_i)$ due to the trivial equation (1.4):

$$C(t_i)=p(t_i)U(t_i) \tag{1.4}$$

We consider the time series of random values $C(t_i)$ and volumes $U(t_i)$ of consecutive trades made with an interval $\varepsilon_c$ during $\Delta$ (1.3) as the origin of price stochasticity. We derive the variance and skewness of price that account for the random volumes of consecutive trades. We propose the equations that determine higher market-based statistical moments of prices. The impact of random volumes of consecutive trades highlights the existing power action of the market trades' randomness on economic processes. The randomness of values and volumes of consecutive trades is the essential property of financial and economic markets that govern the evolution of prices, returns, and macroeconomic variables. We outline market trades made in the economy during a particular averaging period determine the change of macroeconomic variables almost in the similar way as trades made at the exchange determine the mean and variance of prices and returns during $\Delta$ (1.3). We show that the frequency-based statistical moments of prices (1.2) describe a rather limited economic case when all volumes $U(t_i)$ of consecutive trades made during $\Delta$ (1.3) are assumed constant. The implicit use of the limited approximation of constant trade volumes while developing macroeconomic and market models and their forecasts may lead to rather wrong results. To improve the reliability and accuracy of their models and forecasts, the developers of BlackRock's Aladdin, JP Morgan, the U.S. Fed., should reconsider the implicit approximations they use and to account for the essential effects of random market trade.

In Section 2, we consider VWAP as market-based mean price and highlight its reduction to the conventional mean price in case of constant trade volumes. In Section 3, we derive market-based variance of price. In Section 4, we present market-based 3rd statistical moment and Skewness of price. Section 5 describes the limitations of the predictability of statistical moments of price. In Section 6, we discuss the limitations of reliability of Value-at Risk by Gaussian approximations. Section 7 proves that price-volume correlations are always zero. To study price-volume empirical statistical dependence one should calculate correlations between prices and squares of volumes or correlations between squares of prices and volumes. We present the relations that determine these correlations. Conclusion in Section 8. In App. A, we derive market-based variance. In App. B, we derive the 3rd market-based statistical moment and Skewness of price.

We assume that readers are familiar with asset pricing, probability theory, statistical moments, etc., or can find on their own the notions that are not given in the text. All prices are adjusted to current time $t$.

## 2. Market-based mean price

One can equally describe a random variable by its probability measure, characteristic function, and a set of the $n^{th}$ statistical moments (Shephard, 1991; Shiryaev, 1999; Shreve, 2004). In this paper we describe the dependence of statistical moments of price on statistical moments and correlations of the values and volumes of consecutive trades during the averaging interval $\Delta$ (1.3). We underline that market-based statistical moments of price account for the random volumes of consecutive trades during the averaging interval $\Delta$ (1.3). Below we show that frequency-based statistical moments (1.2) describe the limited case of market-based statistical moments when one assumes that all volumes of consecutive trades are constant during the averaging interval. The current assessments of the mean and variance of price during $\Delta$ (1.3) give ground for the predictions of the mean and variance at a horizon $T$. To derive current mean and variance at time $t$, we consider the time series of the values $C(t_i)$, volumes $U(t_i)$, and prices $p(t_i)$ (1.4) of consecutive trades made during the interval $\Delta$ (1.3).

As market-based mean price $p(t;1)$ (2.1) we take the well-known definition of volume weighted average price (VWAP) that was described by (Berkowitz et al., 1988; Buryak and Guo, 2014; Busseti and Boyd, 2015; CME Group, 2020; Duffie and Dworczak, 2021).

$$p(t;1) = \frac{\sum_{i=1}^{N} p(t_i)U(t_i)}{\sum_{i=1}^{N} U(t_i)} = \frac{C_\Sigma(t;1)}{U_\Sigma(t;1)} = \frac{C(t;1)}{U(t;1)} \quad (2.1)$$

The market-based mean price $p(t;1)$ equals the ratio of the total value $C_\Sigma(t;1)$ to total volume $U_\Sigma(t;1)$ (2.2) of consecutive trades made during $\Delta$ (1.3). The ratio of mean value $C(t;1)$ to mean volume $U(t;1)$ (2.2) of trades gives another expression of VWAP mean price $p(t;1)$ (2.1).

$$C_\Sigma(t;1) = \sum_{i=1}^{N} C(t_i) = N \cdot C(t;1) \quad ; \quad U_\Sigma(t;1) = \sum_{i=1}^{N} U(t_i) = N \cdot U(t;1) \quad (2.2)$$

If one assumes that all volumes $U(t_i)$ of trades during $\Delta$ (1.3) are constant and $U(t_i)=U$, then VWAP $p(t;1)$ (2.1) takes the form of the frequency-based mean price $\pi(t;1)$ (1.2):

$$p(t;1)_{|U(t_i)-const} = \frac{1}{N \cdot U} \cdot \sum_{i=1}^{N} p(t_i) \cdot U = \frac{1}{N} \cdot \sum_{i=1}^{N} p(t_i) = \pi(t;1) \quad (2.3)$$

We highlight that one may consider VWAP $p(t;1)$ (2.1) as averaging over the weight function $w(t_i;1)$ (2.4):

$$w(t_i;1) = \frac{U(t_i)}{\sum_{i=1}^{N} U(t_i)} \quad ; \quad \sum_{i=1}^{N} w(t_i;1) = 1 \quad ; \quad p(t;1) = \sum_{i=1}^{N} p(t_i) \cdot w(t_i;1) \quad (2.4)$$

The weight function $w(t_i;1)$ (2.4) determines only the 1st market-based statistical moment of price $p(t;1)$ (2.1; 2.4) and doesn't have the meaning of probability measure. Market-based mean price $p(t;1)$ (2.1) is the consequence of the price equation (1.4). The n-th power of price $p^n(t_i)$ is determined by the equation (2.5):

$$C^n(t_i) = p^n(t_i)U^n(t_i) \quad ; \quad n = 1,2,3,.. \quad (2.5)$$

Similar to the weight function $w(t_i;1)$ (2.4) that is determined by the equation (1.4), we introduce the weight functions $w(t_i;n)$ (2.6) that are determined by the equations (2.5):

$$w(t_i;n) = \frac{U^n(t_i)}{\sum_{i=1}^{N} U^n(t_i)} \quad ; \quad \sum_{i=1}^{N} w(t_i;n) = 1 \quad ; \quad n = 1,2,.. \quad (2.6)$$

As we show below, the use of the weight functions $w(t_i;n)$ (2.6) is very handy for the derivation of statistical moments of price.

The equations (2.5) reveal that the *n-th* statistical moment of price $p(t;n)=E_m[p^n(t_i)]$ should depend on the *n-th* statistical moments of values $C(t;n)$ (2.7) and volumes $U(t;n)$ (2.8) and their mutual averages (2.9) and covariances (2.10):

$$E[C^n(t_i)] = C(t;n) = \frac{1}{N}\sum_{i=1}^{N} C^n(t_i) = \frac{1}{N} C_\Sigma(t;n) \quad ; \quad n = 1,2,\ldots \quad (2.7)$$

$$E[U^n(t_i)] = U(t;n) = \frac{1}{N}\sum_{i=1}^{N} U^n(t_i) = \frac{1}{N} U_\Sigma(t;n) \quad (2.8)$$

$$E[C^n(t_i)U^m(t_i)] = \frac{1}{N}\sum_{i=1}^{N} C^n(t_i)U^m(t_i) \quad (2.9)$$

$$cov[C^n(t_i), U^m(t_i)] = \frac{1}{N}\sum_{i=1}^{N}[C^n(t_i) - C(t;n)][U^m(t_i) - U(t;m)] \quad (2.10)$$

We use the notion of market-based mathematical expectation $E_m[...]$ to highlight that the statistical moments of price $p(t;n)=E_m[p^n(t_i)]$ account for the random volumes of consecutive trades and depend on statistical moments of values $C(t;n)$ (2.7) and volumes $U(t;n)$ (2.8).

We denote $E\,[..]$ (2.7; 2.8) as the conventional mathematical expectation that is approximates the *n-th* statistical moments of values $C(t;n)$ (2.7) and volumes $U(t;n)$ (2.8) with $N$ terms of time series during $\Delta$ (1.3). The functions $C_\Sigma(t;n)$ (2.7) and $U_\Sigma(t;n)$ (2.8) denote the sum of the *n-th* power of values and volumes during $\Delta$ (1.3).

The equations (2.5) and the n-th statistical moments of values $C(t;n)$ (2.7) and volumes $U(t;n)$ (2.8) for n=1,2,.. demonstrates that the conventional frequency-based statistical moments of price $\pi(t;n)$ (1.2) describe a limited approximation of constant trade volumes. Indeed, if trade volumes $U(t_i)=U$ are const during $\Delta$ (1.3), one may easily use the equation (2.5) and present the *n-th* statistical moment $C(t;n)$ (2.7) of the values of trades as follows:

$$C(t;n) = \frac{1}{N}\sum_{i=1}^{N} C^n(t_i) = \frac{1}{N}\sum_{i=1}^{N} p^n(t_i)\cdot U^n = U^n \cdot \frac{1}{N}\sum_{i=1}^{N} p^n(t_i) = U^n \cdot \pi(t;n) \quad (2.11)$$

The equation (2.11) proves that the conventional frequency-based statistical moments of price $\pi(t;n)$ (1.2) are the consequence of the equation (2.5) and the *n-th* statistical moment $C(t;n)$ (2.7) of the values of trades for the special limited case when all trade volumes $U(t_i)=U$ are constant during the averaging interval $\Delta$ (1.3). However, the real financial markets demonstrate highly irregular or random time series of the volumes of consecutive trades. The use of the conventional statistical moments of price $\pi(t;n)$ (1.2; 2.11) implicitly means the use of a limited case of constant trade volumes. The use of constat trade volumes approximation for the description of random market trades, prices, and returns is, in some sense, alike to the use of constant steps to model and forecast random Brownian walks. Both results will be very low.

## 3. Market-based variance of price

VWAP $p(t;1)$ (2.1) determines market-based 1st statistical moment. The square of price $p^2(t_i)$ is determined by the equation (2.5) for $n=2$, which is similar to the equation (1.4). The market-based 2nd statistical moment of price $p(t;2)= E_m[p^2(t_i)]$ (3.1) should depend on 2nd statistical moments of the values $C(t;2)$ (2.7) and volumes $U(t;2)$ (2.8) for $n=2$. The 2nd statistical moment of price $p(t;2)$ should be consistent with $p(t;1)$ (2.1) and hence should depend on it. We consider the equation (3.1) as one that describes a such dependence of $p(t;2)$ on $p(t;1)$:

$$p(t;2) = E_m[p^2(t_i)] = E_m[p(t_i)p(t_i)] = E_m{}^2[p(t_i)] + cov[p(t_i), p(t_i)] \quad (3.1)$$

The equation (3.1) takes a simple form (3.2):

$$E_m[p^2(t_i)] = p(t;2) = p^2(t;1) + \Phi(t;1) \tag{3.2}$$

To fulfil the equations (3.1; 3.2) one should define market-based variance $\Phi(t;1)$ (3.3) of price:

$$\Phi(t;1) = cov[p(t_i), p(t_i)] = E_m\left[\left(p(t_i) - p(t;1)\right)^2\right] = p(t;2) - p^2(t;1) \tag{3.3}$$

We calculate the variance $\Phi(t;1)$ (3.3) by the averaging over the weight function $w(t_i;2)$ (2.5):

$$\Phi(t;1) = E_m\left[\left(p(t_i) - p(t;1)\right)^2\right] = \sum_{i=1}^{N}\left(p(t_i) - p(t;1)\right)^2 w(t_i;2) \tag{3.4}$$

We point to the similarity between equations (1.4) and (2.5) and between $p(t;1)$ (2.1; 2.4) and $\Phi(t;1)$ (3.4). The calculation of the variance $\Phi(t;1)$ (3.4) defines the 2nd market-based statistical moment of price $p(t;2)$ (3.5) that is consistent with $p(t;1)$ (2.1; 2.4):

$$p(t;2) = \Phi(t;1) + p^2(t;1) \tag{3.5}$$

We give step-by-step derivations of the variance $\Phi(t;1)$ (3.4) in App. A., (A.8). The market-based variance $\Phi(t;1)$ of price takes the form:

$$\Phi(t;1) = \frac{\psi^2(t) - 2\,\varphi(t) + \chi^2(t)}{1 + \chi^2(t)} \cdot p^2(t;1) \tag{3.6}$$

We present the definitions of coefficient of variation $\psi(C)$ of the values $C(t_i)$, of the coefficient of variation $\chi(U)$ of the volumes $U(t_i)$, and of their covariance $\varphi(C,U)$ in (A.5).

From (3.5) and (3.6), obtain market-based 2nd statistical moment of price $p(t;2)$:

$$p(t;2) = \left[1 + \frac{\psi^2(C) - 2\,\varphi(C,U) + \chi^2(U)}{1 + \chi^2(U)}\right] \cdot p^2(t;1) \tag{3.7}$$

## 4. The 3rd market-based statistical moment

We propose that the requirement that each next market-based statistical moment of price should depend on the previous ones and their covariances may obey for all market-based statistical moments of price. The 2nd statistical moment $p(t;2)$ (3.5) depends on the 1st one $p(t;1)$ and the variance $\Phi(t;1)$ (3.6). The 3rd statistical moment $p(t;3)$ should depend on the 1st $p(t;1)$ and the 2nd $p(t;2)$, and on their covariance, etc. We assume that a such iterative procedure may determine the dependence of the *n-th* statistical moment on the first *(n-1)* statistical moments:

$$p(t;n) = E_m[p^n(t_i)] = E_m[p(t_i)p^{n-1}(t_i)] = E_m[p(t_i)]E_m[p^{n-1}(t_i)] + cov[p(t_i), p^{n-1}(t_i)] \tag{4.1}$$

The statistical moments $p(t;1)=E_m[p(t_i)]$ and $p(t;n-1)=E_m[p^{n-1}(t_i)]$ are already known. To define $p(t;n)$ one should calculate the covariance between prices $p(t_i)$ and their power $p^{n-1}(t_i)$ by the averaging over the weight function $w(t_i;n)$ (2.6):

$$cov[p(t_i), p^{n-1}(t_i)] = \sum_{i=1}^{N}\left(p(t_i) - p(t;1)\right) \cdot \left(p^{n-1}(t_i) - p(t;n-1)\right) \cdot w(t_i;n) \tag{4.2}$$

According to (4.1; 4.2), to derive the 3rd market-based statistical moment $p(t;3)$ one should follow the relations (4.3; 4.4):

$$E_m[p^3(t_i)] = E_m[p(t_i)p^2(t_i)] = E_m[p(t_i)]E_m[p^2(t_i)] + cov[p(t_i), p^2(t_i)] \quad (4.3)$$

We already derived the 1st $p(t;1)$ (2.1) and the 2nd $p(t;2)$ (3.7) statistical moments. Thus, to define p(t;3) (4.4) one should calculate the covariance $cov[p(t_i),p^2(t_i)]$:

$$p(t;3) = p(t;1)p(t;2) + cov[p(t_i), p^2(t_i)] \quad (4.4)$$

To calculate the covariance $cov[p(t_i), p^2(t_i)]$ (4.4) between prices $p(t_i)$ and squares of prices $p^2(t_i)$ one should average it over the weight function $w(t_i;3)$ (2.6). We present the derivation of the covariance $cov[p(t_i), p^2(t_i)]$ (4.4) in (B.12), the 3rd statistical moment $p(t;3)$ (4.4) in (B.15), and market-based Skewness $Sk_m(p)$ (B.19) in App.B.

The derivation of the 4th statistical moment $p(t;4)$ and market-based Kurtosis $Ku_m(p)$ should follow (4.1; 4.2) for $n=4$ and the weight function $w(t_i;4)$ (2.6). The calculations of higher market-based statistical moments follow the same procedures (4.1; 4.2). We omit these rather long calculations.

## 5. The limitations of the predictability of price statistical moments

The forecasting of random price implies the predictions of its probability. The more price statistical moments may be predicted, the more accurate would be the forecasts of the probability. The dependence of market-based mean $p(t;1)$ (2.1), variance $\Phi(t;1)$ (3.6), Skewness $Sk_m(p)$ (B.19) of price on statistical moments and covariances of the values and volumes of market trades ties up the predictions of the first $n$ statistical moments of price with the forecasts of the first $n$ statistical moments and covariances of the values and volumes of consecutive trades at the same horizon $T$.

To forecast the first $n$ statistical moments and covariances of the values and volumes of consecutive trades with shares of a security $A$ at the horizon $T$ during the averaging interval of the same duration as $\Delta$ (1.3), one should predict the market and economic environment that impact on the evolution of trade statistical moments. To a large extend, the predictions of the first $n$ statistical moments of consecutive trades with a security $A$ requires forecasts of the first $n$ statistical moments and covariances of trades with other securities at the exchange, with the market portfolio at the exchange. The forecasts of statistical moments of trades with market portfolio depend on the predictions of statistical moments of trades at other markets, OTC, consumption, any trades in the economy, and on macroeconomic variables that determine the environment of market trades.

Actually, macroeconomic variables are determined as the sums of means of the values or volumes of trades at different markets or as ratios of such sums (Olkhov, 2023a; 2024). However, the forecasts of the means or 1st statistical moments may predict only the mean prices

*p(t;1)* (2.1). To predict the variance *Φ(t;1)* (3.6) or the 2nd statistical moment *p(t;2)* (3.7), one should have the forecasts of the 2nd statistical moments and covariances of values and volumes of trades in all markets of the economy. And that is almost impossible. Modern econometrics utilize comprehensive methodologies (Fox et al., 2025) to estimate macroeconomic variables that have economic sense of the sums of means of the values or volumes of trades in the economy. No 2nd statistical moments or variances of macroeconomic variables, like GDP, production, consumption, investment, etc., are calculated or studied in econometrics (Fox et al., 2025) and macroeconomic models as well. The variances of prices and returns are almost the only variables that depend on the 2nd statistical moments and covariances of values and volumes of market trades that are accounted for in macroeconomic models.

Ultimately, the lack of econometric assessments and macroeconomic relations that consider the 2nd statistical moments and covariances of market trades in the economy results in lack of economic foundations for their reliable and accurate predictions. Without the predictions of the 2nd statistical moments and covariances of market trades it is impossible to develop market-based forecasts of variances *Φ(t;1)* (3.6) or the 2nd statistical moments *p(t;2)* (3.7) of prices of any securities or commodities. The current forecasts of the variances *Φ(t;1)* (3.6) of prices are almost completely the pure bell art of investors but have almost no economic ground.

Econometric assessments and economic-based predictions of the 3rd, 4th, and higher statistical moments of trade values, volumes, and prices are all the more absent. All that for many years to come will limit the number of predicted price statistical moments by the first two and the accuracy of the forecasts by the accuracy of the Gaussian approximations (Olkhov, 2024).

## 6 The risks of Value-at-Risk

The limitations of the accuracy of predictions of asset price probability determine the reliability of Value-at-Risk (VaR) – one of the most widespread tools to hedge the risks of a price change. The basis for VaR was developed more than 30 years ago (Longerstaey and Spencer, 1996; CreditMetrics™, 1997; Choudhry, 2013). "Value-at-Risk is a measure of the maximum potential change in value of a portfolio of financial instruments with a given probability over a pre-set horizon" (Longerstaey and Spencer, 1996). Despite the progress in VaR performance since then, the core features of VaR remain the same. To assess VaR at horizon *T* one should forecast the integral of the left tail of the probability of prices or returns.

As we show above, the predictions of market-based statistical moments of price depend on the forecasts of statistical moments and correlations of the values and volumes of trades. Hence, VaR as a method to hedge large AUM from risks of price change at horizon *T* depends on the forecasts of the statistical moments and correlations of the values and volumes at the same horizon *T*.

As we discussed above, the economic-based reasons limit the number of predicted statistical moments of the values and volumes of market trades by the first two and the accuracy of predictions of market-based price probabilities by Gaussian approximations. That limit the market-based justification of VaR by Gaussian assessments of the integrals of the left tails of the probabilities of prices or returns.

## 7. Price-volume correlations always zero

The empirical assessments of price-volume correlations were described in numerous papers (Tauchen and Pitts, 1983; Karpoff, 1987; Campbell et al., 1993; Llorente et al., 2001; DeFusco et al., 2017). Actually, the correlations $cov[p(t_i),U(t_i)]$ (7.1 7.2;) of random prices $p(t_i)$ and trade volumes $U(t_i)$ are determined by their joint probability and their mean values. The positive or negative empirical assessments of price-volume correlations (Tauchen and Pitts, 1983; Karpoff, 1987; Campbell et al., 1993; Llorente et al., 2001; DeFusco et al., 2017) are the result of the use of VWAP $p(t;1)$ (2.1) in the assumption that all trade volumes $U(t_i)$ during the averaging interval are constant and the mean price $p(t;1)$ equals $\pi(t;1)$ (1.2; 2.3). However, the assessments of correlation between random trade volumes $U(t_i)$ and random prices $p(t_i)$ are inconsistent with the use of the hypothesis of constant trade volumes $U(t_i)$. If one imagines that trade volumes are constant then price-volume correlations are zero.

The correct definition of $cov[p(t_i),U(t_i)]$ (7.1) should obey:

$$cov[p(t),U(t)] = E_m[(p(t_i) - E_m[p(t_i)]) \cdot (U(t_i) - E_m[U(t_i)])] \tag{7.1}$$

The definition (7.1) causes:

$$cov[p(t),U(t)] = E_m[p(t_i)U(t_i)] - E_m[p(t_i)]E_m[U(t_i)] \tag{7.2}$$

The use of (2.4-2.6; 2.15), give:

$$E_m[U(t_i)] = E[U(t_i)] = \frac{1}{N}\sum_{i=1}^{N} U(t_i)$$

$$C(t;1) = E_m[p(t_i)U(t_i)] = \frac{1}{N}\sum_{i=1}^{N} p(t_i)U(t_i) = \frac{1}{\sum_{i=1}^{N} U(t_i)}\sum_{i=1}^{N} p(t_i)U(t_i) \cdot \frac{1}{N}\sum_{i=1}^{N} U(t_i)$$

Hence, from (2.1; 2.2), obtain:

$$C(t;1) = E_m[p(t_i)]E[U(t_i)] = p(t;1)U(t;1)$$

Thus, the correlation $cov\{p(t),U(t)\}$ (7.2; 7.3) of prices $p(t_i)$ and volumes $U(t_i)$ is always zero:

$$cov\{p(t),U(t)\} = C(t;1) - p(t;1)U(t;1) = 0 \tag{7.3}$$

Actually, the empirical researchers (Tauchen and Pitts, 1983; Karpoff, 1987; Campbell et al., 1993; Llorente et al., 2001; DeFusco et al., 2017) considered the "conventional" definition of price-volume correlation $cov\{p(t),U(t)\}$, which calculates the mean price under the implicit assumption of constant trade volumes:

$$cov\{p(t), U(t)\} = \frac{1}{N}\sum_{i=1}^{N}(p(t_i) - E_m[p(t_i)]) \cdot (U(t_i) - E_m[U(t_i)]) \quad (7.4)$$

The use of (7.4) results in the use of VWAP $p(t;1)$ in the case of constant trade volumes, when $p(t;1)=\pi(t;1)$ (1.2; 2.3):

$$cov\{p(t), U(t)\} = \frac{1}{N}\sum_{i=1}^{N} p(t_i)U(t_i) - U(t;1)\frac{1}{N}\sum_{i=1}^{N} p(t_i) = C(t;1) - \pi(t;1)U(t;1) \quad (7.5)$$

From (7.3; 7.5) obtain, that the empirical researchers calculated not the correlation of random volumes $U(t_i)$ and prices $p(t_i)$ but the difference (7.6) between VWAP $p(t;1)$ (2.1) and its value $\pi(t;1)$ (1.2; 2.3) in the assumption that all trade volumes $U(t_i)$ are constant:

$$cov\{p(t), U(t)\} = C(t;1) - \pi(t;1)U(t;1) = [p(t;1) - \pi(t;1)] \cdot U(t;1) \quad (7.6)$$

The zero price-volume correlation (7.3) doesn't imply that there is no statistical dependence between random prices and volumes. To assess the statistical dependence between random prices and volumes, the researchers should empirically calculate the correlation $cov\{p(t),U^2(t)\}$ between random prices and squares of trade volumes:

$$cov\{p(t), U^2(t)\} = E[p(t_i)U^2(t_i)] - p(t;1)U(t;2) \quad (7.7)$$

The use of (2.1; 2.8; 2.9), give:

$$cov\{p(t), U^2(t)\} = \frac{1}{N}\sum_{i=1}^{N} C(t_i)U(t_i) - p(t;1)U(t;2) \quad (7.8)$$

One may easily derive another form of the same correlation $cov\{p(t),U^2(t)\}$:

$$E[p(t_i)U^2(t_i)] = E[C(t_i)U(t_i)] = C(t;1)U(t;1) + cov\{C(t), U(t)\}$$

From (2.1; A.5-A.7), obtain:

$$cov\{p(t), U^2(t)\} = [\varphi(C,U) - \chi^2(U)]p(t;1)U^2(t;1) \quad (7.9)$$

One may also empirically consider the correlation $cov\{p^2(t),U^2(t)\}$ between squares of random prices and trade volumes:

$$cov\{p^2(t), U^2(t)\} = E[p^2(t_i)U^2(t_i)] - p(t;2)U(t;2) \quad (7.10)$$

The use of (A.5-A.7), give:

$$cov\{p^2(t), U^2(t)\} = \left(1 + \psi^2(C)\right)p^2(t;1)U^2(t;1) - p(t;2)\left(1 + \chi^2(U)\right)U^2(t;1)$$

The use of (B.13; B.14), allows transform the correlation $cov\{p^2(t),U^2(t)\}$ as follows:

$$cov\{p^2(t), U^2(t)\} = 2\,[\varphi(C,U) - \chi^2(U)]p^2(t;1)U^2(t;1) \quad (7.11)$$

From (7.9; 7.11), obtain simple relations between correlation $cov\{p(t),U^2(t)\}$ and correlation $cov\{p^2(t),U^2(t)\}$:

$$cov\{p^2(t), U^2(t)\} = 2cov\{p(t), U^2(t)\}p(t;1) \quad (7.12)$$

The researchers may use (7.8; 7.9; 7.11; 7.12) for empirical investigation of the statistical dependence between random prices and trade volumes.

## 8. Conclusion

The time series of values and volumes of consecutive trades at financial markets and the exchanges are highly irregular or random. Such random dynamics of real trade volumes causes that to derive reliability and accuracy of assessments of current means and variances and of their forecasts one should account for the randomness of market trades. We derive market-based mean, variance and 3rd statistical moment of price that account for the random volumes of consecutive trades made at the exchange with shares of a security. We propose the rules for the derivation of higher market-based statistical moments of prices but don't present the formal proof.

We show that the usual frequency-based statistical moments of price describe only a limited market case when all trade volumes are assumed constant during the averaging interval. The use of frequency-based statistical moments of price for modelling and forecasting of financial markets and macroeconomic environment in some sense is likely the use of constant steps for modelling and predictions of random Brownian walks.

The market-based mean and variance of price depend on statistical moments and covariance of random values and volumes of consecutive trades during the averaging interval. The researchers may control the time periods between the consecutive trades by deriving the sums of all values and volumes of trades made during the selected the period $\varepsilon_c$.

Our results highlight the economic ties between the predictions of market-based mean and variance of prices and the forecasting of statistical moments and covariance of values and volumes of trades at the same horizon during a particular averaging interval. The lack of econometric assessments and macroeconomic relations that consider the 2nd statistical moments and covariances of market trades in the modern economy results in a lack of economic foundations for their reliable and accurate predictions. Without the predictions of the 2nd statistical moments and covariances of market trades, it is impossible to develop market-based forecasts of variances $\Phi(t;1)$ (3.6). That reduces the economically founded forecasts of the statistical moments of price by the first two and the accuracy of predictions of price probability by the accuracy of the Gaussian approximations. Market-based price probability reveals the economic limits on the accuracy of Value-at-Risk. The explicit account for the randomness of the volumes of consecutive trades causes that price-volume correlation to always equal zero. To study price-volume empirical statistical dependence, one should calculate correlations between prices and squares of trade volumes or correlations between squares of prices and squares of trade volumes.

The explicit accounting for the randomness of trade volumes and the use of market-based statistical moments of asset prices may improve the accuracy and reliability of large macroeconomic and market models like BlackRock's Aladdin, JP Morgan, and the U.S. Fed.

## Appendix A. The derivation of market-based variance

To calculate the variance $\Phi(t;1)$ (3.4) we use (2.5 - 2.8) and transform (3.4) as follows :

$$\Phi(t;1) = F(1) + F(2) + p^2(t;1) \tag{A.1}$$

$$F(1) = \sum_{i=1}^{N} p^2(t_i)\, w(t_i;2) = \frac{1}{U(t;2)} \frac{1}{N} \sum_{i=1}^{N} p^2(t_i)\, U^2(t_i) = \frac{C(t;2)}{U(t;2)} \tag{A.2}$$

$$F(2) = -2p(t;1) \cdot \sum_{i=1}^{N} p(t_i)\, w(t_i;2) = \frac{2p(t;1)}{U(t;2)} \frac{1}{N} \sum_{i=1}^{N} p(t_i)\, U^2(t_i)$$

From (2.5) and (2.9; 2.10), obtain:

$$F(2) = -\frac{2p(t;1)}{U(t;2)} \frac{1}{N} \sum_{i=1}^{N} p(t_i)\, U^2(t_i) = \frac{2p(t;1)}{U(t;2)} \cdot \frac{1}{N} \sum_{i=1}^{N} C(t_i)\, U(t_i)$$

$$\frac{1}{N} \sum_{i=1}^{N} C(t_i)\, U(t_i) = E[C(t_i)U(t_i)] = C(t;1)U(t;1) + cov[C(t_i), U(t_i)]$$

The covariance $cov[C(t_i),U(t_i)]$ between values and volumes has the conventional form (2.10):

$$cov[C(t_i), U(t_i)] = \frac{1}{N} \sum_{i=1}^{N} [C(t_i) - C(t;1)]\, [U(t_i) - U(t;1)]$$

Finaly, obtain:

$$F(2) = -\frac{2p(t;1)}{U(t;2)} \left[ C(t;1)U(t;1) + cov[C(t_i), U(t_i)] \right] \tag{A.3}$$

From (A.1-A.3), obtain:

$$\Phi(t;1) = \frac{C(t;2) - 2p(t;1)C(t;1)U(t;1) - 2p(t;1)cov[C(t_i),U(t_i)] + p^2(t;1)U(t;2)}{U(t;2)} \tag{A.4}$$

To transform (A.4) to more easy form let us introduce coefficients of variation $\psi(C)$ (A.5) of the values $C(t_i)$, the coefficient of variation $\chi(U)$ (A.5) of the volumes $U(t_i)$, and their covariance $\varphi(C,U)$ (A.5) normalized to their mean values $C(t;1)$ and volumes $U(t;1)$ (2.2):

$$\psi^2(C) = \frac{cov[C(t),C(t)]}{C^2(t;1)} \quad ; \quad \chi^2(U) = \frac{cov[U(t),U(t)]}{U^2(t;1)} \quad ; \quad \varphi(C,U) = \frac{cov[C(t),U(t)]}{C(t;1)U(t;1)} \tag{A.5}$$

The use of (A.5) presents $C(t;2)$, $U(t;2)$ as (A.6) and the covariance $cov[C(t_i),U(t_i)]$ as (A.7):

$$C(t;2) = \left(1 + \psi^2(C)\right) C^2(t;1) \quad ; \quad U(t;2) = (1 + \chi^2(U))U^2(t;1) \tag{A.6}$$

$$cov[C(t_i), U(t_i)] = \varphi(C,U)C(t;1)U(t;1) \tag{A.7}$$

The substitution of (A.5-A.7) into (A.4), gives:

$$\Phi(t;1) = \frac{\psi^2(C) - 2\,\varphi(C,U) + \chi^2(U)}{1 + \chi^2(U)} \cdot p^2(t;1) \tag{A.8}$$

## Appendix B. The derivation of the 3rd market-based statistical moment

The equation (4.2) for *n=3* determines the market-based covariance $cov[p(t_i),p^2(t_i)]$:

$$cov[p(t_i),p^2(t_i)] = \sum_{i=1}^{N}\big(p(t_i)-p(t;1)\big)\big(p^2(t_i)-p(t;2)\big)\,w(t_i;3) \qquad \text{(B.1)}$$

Similar to (A.1), one may present the averaging of the polynomial in (B.1) as (B.2):

$$cov[p(t_i),p^2(t_i)] = G(1)+G(2)+G(3)+p(t;1)p(t;2) \qquad \text{(B.2)}$$

The functions *G(1), G(2),* and *G(3)* have the following forms (2.7-2.10):

$$G(1) = \sum_{i=1}^{N} p^3(t_i)\,w(t_i;3) = \frac{1}{U(t;3)}\cdot\frac{1}{N}\sum_{i=1}^{N} p^3(t_i)\,U^3(t_i) = \frac{C(t;3)}{U(t;3)} \qquad \text{(B.3)}$$

$$G(2) = -\frac{p(t;1)}{U(t;3)}\cdot\frac{1}{N}\sum_{i=1}^{N} p^2(t_i)\,U^3(t_i) = -\frac{p(t;1)}{U(t;3)}\frac{1}{N}\sum_{i=1}^{N} C^2(t_i)\,U(t_i) \qquad \text{(B.4)}$$

$$G(3) = -\frac{p(t;2)}{U(t;3)}\frac{1}{N}\sum_{i=1}^{N} C(t_i)\,U^2(t_i) \qquad \text{(B.5)}$$

The relations (2.9; 2.10) the covariance $cov[C^2(t_i),U(t_i)]$ as follows:

$$\frac{1}{N}\sum_{i=1}^{N} C^2(t_i)\,U(t_i) = E[C^2(t_i)U(t_i)] = E[C^2(t_i)]E[U(t_i)] + cov[C^2(t_i),U(t_i)]$$

$$\frac{1}{N}\sum_{i=1}^{N} C^2(t_i)\,U(t_i) = C(t;2)U(t;1) + cov[C^2(t_i),U(t_i)] \qquad \text{(B.6)}$$

Thus, the function *G(2)* takes the form (B.7):

$$G(2) = -\frac{p(t;1)}{U(t;3)}\,\big[C(t;2)U(t;1) + cov[C^2(t_i),U(t_i)]\big] \qquad \text{(B.7)}$$

Similar to G(2) (B.7), from (B.5) obtain expression for G*(3)*:

$$G(3) = -\frac{p(t;2)}{U(t;3)}\,\big[C(t;1)U(t;2) + cov[C(t_i),U^2(t_i)]\big] \qquad \text{(B.8)}$$

Similar to (A.5), we denote coefficients of variation $\psi(C,C^2)$ (B.9) of values and coefficients of variation $\chi(U,U^2)$ (B.9) of volumes:

$$\psi(C,C^2) = \frac{cov[C(t_i),C^2(t_i)]}{C(t;1)C(t;2)} \quad ; \quad \chi(U,U^2) = \frac{cov[U(t_i),U^2(t_i)]}{U(t;1)U(t;2)} \qquad \text{(B.9)}$$

The coefficients of covariances $\varphi(C^2,U)$ (B.10) and $\varphi(C,U^2)$ (B.11) take the form:

$$\varphi(C^2,U) = \frac{cov[C^2(t_i),U(t_i)]}{C(t;2)U(t;1)} = \frac{cov[C^2(t_i),U(t_i)]}{\big(1+\psi^2(C)\big)C^2(t;1)U(t;1)} \qquad \text{(B.10)}$$

$$\varphi(C,U^2) = \frac{cov[C(t_i),U^2(t_i)]}{C(t;1)U(t;2)} = \frac{cov[C(t_i),U^2(t_i)]}{\big(1+\chi^2(U)\big)C(t;1)U^2(t;1)} \qquad \text{(B.11)}$$

Simple but long transformations give:

$$C(t;3) = E[C^3(t_i)] = E[C(t_i)C^2(t_i)] = C(t;1)C(t;2) + cov[C,C^2]$$

$$C(t;3) = C(t;1)C(t;2)\big(1+\psi(C,C^2)\big)$$

$$\frac{1}{N}\sum_{i=1}^{N} C^2(t_i)\,U(t_i) = E[C^2(t_i)U(t_i)] = C(t;2)U(t;1) + cov[C^2(t_i),U(t_i)]$$

$$\frac{1}{N}\sum_{i=1}^{N} C^2(t_i)\,U(t_i) = C(t;2)U(t;1)\big(1+\varphi(C^2,U)\big)$$

$$\frac{1}{N}\sum_{i=1}^{N} C(t_i)\,U^2(t_i) = C(t;1)U(t;2)\big(1+\varphi(C,U^2)\big)$$

Finaly, obtain:

$$cov[p(t_i), p^2(t_i)] = G(1) + G(2) + G(3) + p(t;1)p(t;2) =$$

$$= [\frac{(1+\psi^2(C))(\psi(C,C^2)-\varphi(C^2,U))-(1+\varphi(C,U^2))(1+\chi^2(U))P(2)}{(1+\chi^2(U))(1+\chi(U,U^2))} + P(2)] \cdot p^3(t;1) \quad \text{(B.12)}$$

$$p(t;2) = \left[1 + \frac{\psi^2(C)-2\,\varphi(C,U)+\chi^2(U)}{1+\chi^2(U)}\right] \cdot p^2(t;1) = P(2) \cdot p^2(t;1) \quad \text{(B.13)}$$

$$P(2) = \left[1 + \frac{\psi^2(C)-2\,\varphi(C,U)+\chi^2(U)}{1+\chi^2(U)}\right] \quad \text{(B.14)}$$

$$p(t;3) = \frac{(1+\psi^2(C))(\psi(C,C^2)-\varphi(C^2,U))+(1+2\chi(U,U^2)-\varphi(C,U^2))(1+\chi^2(U))P(2)}{(1+\chi^2(U))(1+\chi(U,U^2))}\, p^3(t;1) \quad \text{(B.15)}$$

One may checkup that if all trade volumes *U(t<sub>i</sub>)=U* are constant during *Δ* (1.3), then:

$$\chi^2(U) = \chi(U,U^2) = \varphi(C,U^2) = \varphi(C^2,U) = 0$$

$$p(t;3) = \frac{(1+\psi^2(C))\psi(C,C^2) + P(2)}{1}\, p^3(t;1) = [(1+\psi^2(C))\psi(C,C^2) + P(2)]p^3(t;1)$$

$$P(2) = \left[1 + \frac{\psi^2(C)-2\,\varphi(C,U)+\chi^2(U)}{1+\chi^2(U)}\right] = 1 + \psi^2(C) \quad \text{(B.16)}$$

$$p(t;3)_{|U-const} = (1+\psi^2(C))_{|U-const}(1+\psi(C,C^2)_{|U-const}p^3(t;1)_{|U-const} \quad \text{(B.17)}$$

From (2.3; 2.11), obtain:

$$1+\psi^2(C)_{|U-const} = \frac{C^2(t;1)+cov[C(t),C(t)]}{C^2(t;1)}_{|U-const} = \frac{\pi(t;2)}{\pi^2(t;1)}$$

$$(1+\psi(C,C^2)_{|U-const} = \frac{C(t;1)C(t;2)+cov[C(t),C^2(t)]}{C(t;1)C(t;2)}_{|U-const} = \frac{\pi(t;3)}{\pi(t;1)\pi(t;2)}$$

Finally, obtain for (B.17), as it should be:

$$p(t;3)_{|U-const} = \pi(t;3) \quad \text{(B.18)}$$

One may use the above results to obtain the market-based Skewness $Sk_m(p)$ (B.19):

$$Sk_m(p)\Phi^{3/2}(t;1) = E_m\left[(p(t_i) - p(t;1))^3\right] = p(t;3) - 3p(t;2)p(t;1) + 2p^3(t;1)$$

$$Sk_m(p)\Phi^{\frac{3}{2}}(t;1) =$$

$$= \left\{2 + \frac{(1+\psi^2(C))(\psi(C,C^2)-\varphi(C^2,U))-(2+\chi(U,U^2)+\varphi(C,U^2))(1+\chi^2(U))P(2)}{(1+\chi^2(U))(1+\chi(U,U^2))}\right\} p^3(t;1) \quad \text{(B.19)}$$